\documentclass[aps,prb,showpacs]{revtex4}%
\usepackage{amsfonts}
\usepackage{amsmath}
\usepackage{amssymb}
\usepackage{graphicx}%
\providecommand{\U}[1]{\protect \rule{.1in}{.1in}}
\providecommand{\U}[1]{\protect \rule{.1in}{.1in}}
\begin{document}
\title{Exchange effects on electron transport through single-electron spin-valve transistors}
\author{Wouter Wetzels}
\author{Gerrit E.W. Bauer}
\affiliation{Kavli Institute of Nanoscience, Delft University of Technology, Lorentzweg 1,
2628 CJ Delft, The Netherlands}
\author{Milena Grifoni}
\affiliation{Institut f\"{u}r Theoretische Physik, Universit\"{a}t Regensburg, 93035
Regensburg, Germany}


\begin{abstract}
We study electron transport through single-electron spin-valve transistors in
the presence of non-local exchange between the ferromagnetic leads and the
central normal-metal island. The Coulomb interaction is described with the
\textquotedblleft orthodox model\textquotedblright \ for Coulomb blockade and
we allow for noncollinear lead magnetization directions. Two distinct
exchange mechanisms that have been discussed in the literature are shown to be of
comparable strength and are taken into account on equal footing. We present
results for the linear conductance as a function of gate voltage and magnetic
configuration, and discuss the response of the system to applied magnetic fields.

\end{abstract}
\maketitle

\section{Introduction}

Downscaling magnetoelectronic devices to the nanometer regime implies that
electron-electron interaction effects become prominent, as has been amply
demonstrated by many experimental studies on the Coulomb blockade in double
tunnel junctions with ferromagnetic
elements.\cite{Ono,Schelp,Ralph,Pasupathy,Philip,Zhang,Sahoo,Yakushiji,Bernand-Mantel}
Much of the theoretical work focusses on F$|$N$|$F spin valves, in which the
island is a normal metal (N) and the contacts are ferromagnets (F) with
variable magnetization directions. Initially, the interest was mainly focussed
on the giant magnetoresistance, \textit{i.e}. the difference in the transport
properties for parallel or antiparallel magnetizations.\cite{Barnas, Majumdar,
Korotkov, Brataasnib, Brataas and Wang} More recently, the interplay between
spin and interaction effects for non-collinear magnetization configurations
has attracted quite some
interest.\cite{Balents,Bena,Koenig,Gorelik,Rudzinski,Pedersen,Fransson,Braig,Wetzels,Parcollet,Mu}%

A single-electron spin-valve transistor (SV-SET) is an F$|$N$|$F spin-valve
with a sufficiently small normal-metal (N) island that is coupled to the
ferromagnetic leads by tunnel barriers. When the electrostatic charging energy
of the island is larger than the thermal energy, charge transport can be
controlled at the level of single electron charges by varying bias and gate
voltage, as is well known for non-magnetic SET's.\cite{Grabert} With
spin-dependent electron tunneling rates and sufficiently long spin-decay
lifetimes, a spin accumulation (or non-equilibrium magnetization) that strongly 
affects electron transport may build up in the nonmagnetic island.

In this article, we discuss the transport characteristics of metallic SV-SETs
in the Coulomb blockade regime, allowing for arbitrary, noncollinear
magnetization directions. In particular, we examine the influence of exchange
effects through F$|$N tunnel contacts on the spin accumulation in the center
island, presenting a more complete discussion compared to that in Ref.
\onlinecite{Wetzels}. We argue that two separate exchange effects have to be
taken into account. On one hand, there is the non-local interface exchange,
let us call it \textquotedblleft X1\textquotedblright \ in the following. In
scattering theory for non-interacting systems it is described by the imaginary part
of the spin-mixing conductance,\cite{Brataas} while in the context of
current-induced magnetization dynamics X1 acts as an \textquotedblleft effective
field\textquotedblright.\cite{Stiles} Such an effective field has been found
experimentally to strongly affect the transport dynamics in spin valves with
MgO tunnel junctions.\cite{Tulapurkar} This effect has recently also been
involved to explain magnetoresistance effects in carbon nanotube spin
valves\cite{Sahoo} and called spin-dependent interface phase
shifts.\cite{Cottet} The second exchange term (\textquotedblleft
X2\textquotedblright) is an interaction-dependent exchange effect due to
virtual tunneling processes that is absent in non-interacting systems. It has
been considered for islands in the electric quantum limit, in which transport
is carried by a single quantized level only.\cite{Koenig} The X2 effect is
potentially attractive for quantum information processing, since it allows to
switch on and off effective magnetic fields in arbitrary directions just by a
gate electric potential. We compute here X2 for a metallic island in which
size quantization is not important. We find that both exchange effects are of
comparable magnitude and affect the transport properties in a characteristic
way, but can be separated in principle by employing the gate dependence of X2.

The paper is organized as follows. In Sec. \ref{model} we introduce the model
system for the SV-SET. In Sec. \ref{exchange} the two relevant types of
exchange processes are discussed. Charge and spin transfer rates are
determined in Sec. \ref{transport}. Finally, we present results for the
transport characteristics as a function of magnetic configuration, gate
voltage and applied magnetic field in Sec. \ref{results}.

\section{Model system}

\label{model} 
\begin{figure}[tbh]
\centering \includegraphics[width=3.375in]{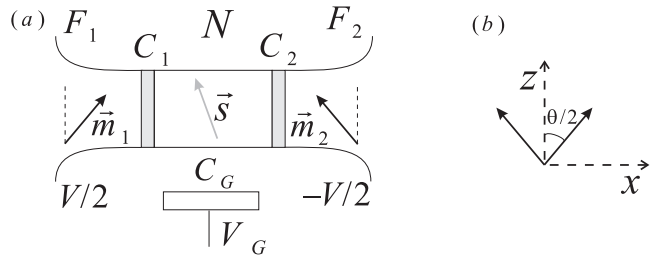}\caption{(a) The
spin-valve single-electron transistor: A small normal-metal island
tunnel-coupled to two large ferromagnetic leads. The unpaired spin angular
momentum on the island is denoted by $\vec{s}$. (b) The magnetization
directions in the leads define an angle $\theta$.}%
\label{schematic}%
\end{figure}

An SV-SET (see Fig. \ref{schematic}a) is composed of a small
metallic cluster in contact with two large ferromagnetic electron reservoirs
in thermal equilibrium characterized by magnetization directions
$\overrightarrow{m}_{1}$ and $\overrightarrow{m}_{2}$ with $\overrightarrow
{m}_{\alpha}=\left(  \sin \theta_{\alpha},0,\cos \theta_{\alpha}\right)  $ (for
$\alpha=1,2$), where $\theta_{1}=\theta/2$ and $\theta_{2}=-\theta/2$.

The F$_{\alpha}|$N contacts are tunneling barriers with
conductances that depend on the electron spin, $G_{\alpha}^{\uparrow \uparrow}$
for the majority and $G_{\alpha}^{\downarrow \downarrow}$ for the minority spin
in the ferromagnet. The total conductance for contact $\alpha$ is then given
by $G_{\alpha}\equiv \left(  G_{\alpha}^{\uparrow \uparrow}+G_{\alpha
}^{\downarrow \downarrow}\right)  $ and the contact polarization is defined as
$P_{\alpha}\equiv \left(  G_{\alpha}^{\uparrow \uparrow}-G_{\alpha}%
^{\downarrow \downarrow}\right)  /\left(  G_{\alpha}^{\uparrow \uparrow
}+G_{\alpha}^{\downarrow \downarrow}\right)  $. The resistances $R_{\alpha
}=1/G_{\alpha}$ are taken to be much larger than the resistance quantum
$R_{Q}=h/e^{2}$, which, at low enough temperatures and bias voltages, allows
us to study the\ blockade of transport by the Coulomb interaction. The
electron tunneling rates are governed by the change of electrostatic energy of
the whole circuit upon transfer of an electron. The capacitances of the
junctions $C_{\alpha}$ determine the charging energy of the island.

We limit our considerations to islands formed by metallic clusters for which 
the thermal energy ($k_{B}T$) is much larger than the average single-particle energy
separation (reciprocal density of states) $\delta=1/\rho_{N}$, but much
smaller than the single-electron charging energy. Therefore, many levels on
the island participate in the transport and we may treat the electronic
spectrum as continuous. For a gold cluster with a diameter of 10 $\operatorname{nm}$, $\delta$ approximately corresponds to a temperature of 2 K.  The Kondo physics of quantum dots connected to
ferromagnetic leads\cite{ZhangXue,Martinek,Lopez,Pasupathy} is suppressed in
this regime.

Since the currents flowing into and out of the cluster are spin-polarized, the
island may become magnetized. The number of unpaired spins on the island is
limited by spin-flip scattering, which we parametrize by a spin-flip
relaxation time $\tau_{\mathrm{sf}}$. There is evidence from several
experiments that the spin-flip times in metallic nanoparticles can be much
longer than in bulk systems, which implies that the effects of a spin
accumulation on the island should be taken into account.\cite{Yakushiji,
Bernand-Mantel, Zhang} For later convenience we introduce the spin-flip
conductance parameter $G_{sf}\equiv \rho_{N}e^{2}/\left(  2\tau_{\mathrm{sf}%
}\right)  $. We assume that the energy relaxation on the island is fast
compared to the electron dwell time.

The total Hamiltonian for the SV-SET is
\begin{equation}
H=H_{N}+\sum \limits_{\alpha=1,2}\left(  H_{F\alpha}+H_{T\alpha}+H_{\mathrm{ex}%
\alpha}\right)  ,
\end{equation}
where $H_{N}$ is the Hamiltonian for the normal metal cluster in the
\textquotedblleft orthodox model\textquotedblright \ for Coulomb blockade
\begin{equation}
H_{N}=\sum \limits_{ks}\varepsilon_{k}c_{ks}^{\dagger}c_{ks}+\frac{e^{2}\left(
n_{N}-C_{G}V_{G}/e\right)  ^{2}}{2C}.
\label{HN}
\end{equation}
Here $c_{ks}^{\dagger}$ is a creation operator for an electron state with orbital
index $k$ and spin $s\in \{ \uparrow,\downarrow \},$ where the $z$-axis is chosen
as spin quantization axis. The Hamiltonian includes an electrostatic
interaction energy which depends on the junction capacitances $C_{\alpha}$,
the gate voltage $V_{G}$ and the excess number of electrons on the island
$n_{N}$. The gate voltage shifts the potential and induces a charge
$C_{G}V_{G}$. We assume that the gate capacitance $C_{G}\ll C_{1},C_{2}$, and
in the following we use $C_{1}=C_{2}=C/2$. The energy levels in the two
ferromagnetic leads (denoted by $\alpha=1,2$) are spin-dependent:
\begin{equation}
H_{F\alpha}=\sum_{ks}\varepsilon_{\alpha ks}a_{\alpha ks}^{\dagger}a_{\alpha
ks}.
\end{equation}
The operators $a_{\alpha ks}^{\dagger}$ create electrons with spin $s$ in the
spin-quantization axis along $\overrightarrow{m}_{\alpha}$.

It is convenient to introduce annihilation operators $c_{\alpha ks}$ for
electrons in the normal metal defined for a quantization axis in the direction
of $\overrightarrow{m}_{\alpha}$. The relation between operators in the two
bases is then $c_{\alpha ks}=\hat{U}_{ss^{\prime}}\left(  \theta_{\alpha
}\right)  c_{ks^{\prime}}$, expressed in terms of the spin
${\frac12}$
rotation matrix
\begin{equation}
\hat{U}\left(  \theta_{\alpha}\right)  =e^{i\sigma_{y}\theta_{\alpha}%
/2}=\left(
\begin{array}
[c]{cc}%
\cos \theta_{\alpha}/2 & \sin \theta_{\alpha}/2\\
-\sin \theta_{\alpha}/2 & \cos \theta_{\alpha}/2
\end{array}
\right)  .
\end{equation}
Then, for each contact, a tunneling Hamiltonian
\begin{equation}
H_{T\alpha}=\sum_{kqs}T_{kqs}^{\alpha}a_{\alpha ks}^{\dagger}c_{\alpha
qs}+h.c.
\end{equation}
describes the coupling to the island. The tunneling coefficients are assumed
to not significantly depend on energy on the scale of the charging energy. We
discuss the exchange contribution, represented by the Hamiltonian
$H_{\mathrm{ex}\alpha}$, in the next section.

\section{Exchange effects}

\label{exchange} Here we discuss two different exchange effects that affect
the electrons in the normal metal island attached to magnetic
contacts. These two flavors arise when the transport properties for an SV-SET
are determined to lowest order in the tunnelling probabilities. 
\subsection{Nonlocal interface exchange (X1)}

\label{nonlocal} The non-local exchange coupling between ferromagnetic films
through a normal metal spacer is an important effect that determines the
ground state of magnetic multilayers (see Ref. \onlinecite{Stilesxc} for a
recent review). Electrons in a normal metal that are reflected at a contact to
a ferromagnet, pick up a phase depending on the electron spin relative to the
magnetization direction. In sufficiently clean and narrow F$|$N$|$F
structures, quantum well states are formed in N whose energy depend on the
magnetic configuration through the spin-dependent phase. By a rotation of the
magnetization directions the energy spectrum and Fermi energy varies, causing
the ground state energy to depend on the relative angle $\theta$. In metallic
multilayers with a suitable spacer thickness, this can lead to an antiparallel
ground state, which displays the celebrated giant magnetoresistance when the
magnetizations are forced into a parallel direction by a magnetic field. Even
when the ground state energies are not significantly affected by the exchange
coupling, configuration-dependent quantized states can still be observed in
transport. This has been shown for high-quality planar tunnel
junctions\cite{Yuasa} as well as spin valves in which the node is formed by
single carbon nanotubes with a quantized energy spectrum.\cite{Sahoo,Man} In
Ref. \onlinecite{Cottet} the effect of interfacial phase shifts on the
magnetoresistance of ballistic quantum wires between ferromagnetic leads was
calculated. The spin-dependent phase shifts give rise to a slightly different
quantization condition, which can spin-split the energy levels. Since we are
here interested in classical islands with a continuous electron spectrum, we
calculate energy shifts for a semiclassical island using the Bohr-Sommerfeld
quantization rule in Appendix \ref{appshifts}.

Here we consider the limit of tunnel junctions between a normal metallic
island and ferromagnetic electrodes. The torques on the ferromagnets are then
very small. The exchange coupling does not significantly disturb the
ferromagnets in this limit, but persists to affect transport. The present
study focusses on the charge transport properties in the limit of small
tunneling matrix elements, thus from the outset excluding resonant tunneling,
co-tunneling or Kondo type physics. The states on the island may be size
quantized, \textit{i.e}. the energy level spacing exceeds the thermal energy
(\textquotedblleft quantum dot\textquotedblright), or, in the opposite limit,
better described by a semicontinuous density of states (\textquotedblleft
classical dot\textquotedblright). Here we concentrate on the latter,
\textit{i.e}. semiclassical, diffuse, or chaotic islands, for which it can be
shown quite generally that equilibrium spin currents are
suppressed.\cite{Waintal} The state of the island is then characterized by a
semiclassical charge and spin distribution function that has to be determined
self-consistently as a function of the junction parameters and the applied
voltages. For non-interacting systems, the spin and charge currents through an
F$|$N interface are determined not only by the conventional conductances
$G_{\alpha}^{\uparrow \uparrow}$ and $G_{\alpha}^{\downarrow \downarrow}$
introduced above, but also by the complex spin-mixing conductances $G_{\alpha
}^{\uparrow \downarrow}$,\cite{Brataas} which are discussed in Sec.
\ref{transport}. \ The real part $\operatorname{Re}G_{\alpha}^{\uparrow
\downarrow}$ is the material parameter that is proportional to the
spin-transfer torque.\cite{Slonczewski,Brataasrev} The imaginary part
$\operatorname{Im}G_{\alpha}^{\uparrow \downarrow}$ reflects the spin-dependent
interface phase shifts and affects the magnetization and spin accumulation
dynamics as an effective exchange magnetic field parallel to the magnetization
direction.\cite{Dani,Stiles,Brataasrev} $\operatorname{Im}G_{\alpha}%
^{\uparrow \downarrow}$ is relatively small for intermetallic
interfaces,\cite{Xia} but is in general comparable in magnitude to the other
conductance parameters.\cite{Dani,Zwierzycki} The non-local interface exchange
has been discussed in similar terms for spin valves consisting of Luttinger
liquids with ferromagnetic contacts.\cite{Balents}

The blocking of transport by the Coulomb charging is usually described by
Fermi's Golden Rule (see below), which employs a probability (squared matrix
elements) and energy conservation. As long as the charging energy is much
smaller than atomic energy scales (like the Fermi energy), \ the junction
parameters such as the interface transparency and spin-mixing conductance are
unaffected and the Coulomb blockade is governed by the energy conservation
criterion only. This implies that the exchange effect can be described by the
$\operatorname{Im}G_{\alpha}^{\uparrow \downarrow}$ of the bare junction.

It remains to parametrize the exchange in the limit of the tunneling
Hamiltonian, \textit{i.e}. to lowest order in the interface transmission. We
show below that this is achieved by adding the following exchange term
$H_{\mathrm{ex}\alpha}$ to the Hamiltonian for the two leads:
\begin{equation}
H_{\mathrm{ex}\alpha}=\sum \limits_{ks}\Delta \epsilon_{\alpha ks}c_{\alpha
ks}^{\dagger}c_{\alpha ks}.\label{Hex}%
\end{equation}
The energy shifts $\Delta \epsilon_{\alpha ks}$, see Eq. (\ref{shifts}), are
proportional to the inverse density of states , but they remain relevant for
small level splitting because the dwell time is inversely proportional to
the average energy level separation or inverse density of states $\delta=\rho
_{N}^{-1}$. This Hamiltonian is an effective Zeeman splitting caused by an
exchange magnetic field in the direction of the magnetization, see Sec.
\ref{spinacc}.

\subsection{Virtual tunneling processes (X2)}

\label{subX2} The interface exchange term X1 is a property of the separate
interfaces and they contribute independently. The second type of exchange (X2)
felt by the spins on the island is a property of the entire device. It
originates from virtual tunneling processes, corresponding to single-electron
transfer from and to the cluster. In the tunneling regime, this process can be
treated and understood in terms of perturbation theory. In the absence of
tunneling, the number of electrons on the island is a good quantum number. The
perturbation by the contact to the electrodes allows mixing in of states\ in
which the number of electrons on the island is changed by unity, at the cost
of the charging energy. In second order perturbation theory this results in an
energy gain represented by a sum over (virtually) excited states in which the
Coulomb energy appears in the denominator and the tunneling probability in the
numerator. When the leads are non-magnetic, these virtual processes correspond
to a quantum correction to the average charge on the central
electrode.\cite{Matveev,Glazman} This effect depends strongly on the applied
gate voltage. When the unperturbed $N+1$ $\left(  N-1\right)  $ particle
ground state is tuned in energy just above the $N$ particle state, the quantum
correction will be large and positive (negative). At the degeneracy point,
perturbation theory breaks down, but the ensuing divergence can be controlled
by taking into account finite temperatures. 

When the tunneling probabilities to the ferromagnetic contacts are spin
dependent, the deviations from the exact quantized charge on the island become
spin-dependent, and therefore lead to a net excess of spins in the ground state
that depends on the configuration of the contact magnetizations. For a
symmetric spin valve it is easy to see that the island ground state
magnetization due to these virtual processes X2 is maximal for parallel
magnetizations and vanishes for antiparallel ones.

The additional exchange affects non-equilibrium electron
transport, in contrast to higher-order so-called co-tunneling processes, to the
same order as the in and out-tunneling processes. For a quantum-dot island
with a single quantized level, K\"{o}nig and Martinek\cite{Koenig} showed that
in the case of \textit{non-collinear} magnetizations the non-equilibrium spins
on an island injected by a finite source-drain voltage are dephased by
precessing around the effective exchange field. This effect was also discussed
for few-level quantum dots.\cite{Braig}  Since X1 discussed in Sec.
\ref{nonlocal} is a material constant, the gate voltage dependence of X2
provides a handle for an experimental discrimination of the two effects. We
derive an expression for the effective X2 exchange field for a classical
SV-SET in Sec. \ref{spinacc}.

\section{Charge and spin transport}

\label{transport} We compute the transport characteristics of the SV-SET in
lowest-order perturbation theory\cite{Mahan} for a diffusive or chaotic island
in the sequential tunneling regime. The rate equations lead to a probability
distribution for the excess number of charges $n_{N}$. The excess spin
accumulation $\vec{s}$ contains a large number of spins and we are interested
in its average value in the steady state that is found from the condition
$\langle d\vec{s}/dt\rangle=0$.

\subsection{Charge transfer}

The operators for the excess number of electrons on the island and on the two
leads are $n_{N}=\sum_{ks}c_{ks}^{\dagger}c_{ks}$ and $n_{F\alpha}=\sum
_{ks}a_{\alpha ks}^{\dagger}a_{\alpha ks},$ respectively. The unpaired spin
angular momentum on the cluster is written as $\vec{s}=(\hbar/2)\sum
_{kss^{\prime}}c_{ks}^{\dagger}\vec{\sigma}_{ss^{\prime}}c_{ks^{\prime}}$,
where $\vec{\sigma}=\left(  \sigma_{x},\sigma_{y},\sigma_{z}\right)  $ is the
vector of Pauli spin matrices. It is convenient to introduce a vector chemical
potential $\overrightarrow{\Delta \mu}$ in the island, with size $\left \vert
\overrightarrow{\Delta \mu}\right \vert =2\left \vert \langle \vec{s}%
\rangle \right \vert /\left(  \rho_{N}\hbar \right)  $ \ (see also Ref.
\onlinecite{Brataasnib}), where $\rho_{N}$ is the density of states at the
Fermi energy. We can take into account Stoner enhancement intra-island
exchange effects in terms of the static susceptibility $\chi_{s},$ and we may
also write $\Delta \mu=2\mu_{B}^{2}\left \vert \langle \vec{s}\rangle \right \vert
/\left(  \chi_{s}\hbar \right)$. We denote the unit vector in the direction
of the spin accumulation by $\hat{s}$.

The charge current is equal to the expectation values for the rate of change
of $n_{N}$. In terms of the tunneling Hamiltonian
$H_{T}=H_{T1}+H_{T2}$ the time-evolution is given by
\begin{align}
\frac{dn_{N}}{dt}  &  =\frac{i}{\hbar}\left[  H_{T},n_{N}\right] \nonumber \\
&  =\frac{i}{\hbar}\sum_{\alpha kqs^{\prime}}T_{kqs^{\prime}}^{\alpha
}a_{\alpha ks^{\prime}}^{\dagger}c_{\alpha qs^{\prime}}+h.c.,
\end{align}
We use the interaction representation, and write the total Hamiltonian as $H=H^{\prime}+H_{T}.$
To second order in $H_{T}$ we have
\[
\left \langle \frac{dn_{N}\left(  t\right)  }{dt}\right \rangle =\frac{i}{\hbar
}\int \limits_{-\infty}^{t}dt^{\prime}\left \langle \left[  \frac{dn_{N}\left(
t\right)  }{dt},H_{T}\left(  t^{\prime}\right)  \right]  \right \rangle
_{\circ},
\]
where $\langle..\rangle_{\circ}$ denotes an expectation value with respect to 
Hamiltonian $H^{\prime}$. The electrochemical potentials of the two reservoirs are
$\mu_{cF1}=eV/2$ and $\mu_{cF2}=-eV/2$. It is convenient to introduce grand
canonical Hamiltonians including the chemical potentials
as\cite{Balents,Mahan}
\begin{align}
K_{N}  &  =H_{N}-\hbar^{-1}\overrightarrow{\Delta \mu}\cdot \vec{s},\\
K_{F_{\alpha}}  &  =H_{F_{\alpha}}-\mu_{cF_{\alpha}}n_{F_{\alpha}}.
\end{align}
The time dependence $c_{ks}\left(  t\right)  =e^{\frac{i}{\hbar}K_{N}t}%
c_{ks}e^{-\frac{i}{\hbar}K_{N}t}$ can be formulated in terms of the projection
operators
\begin{align}
\hat{u}^{\uparrow}\left(  \hat{s}\right)   &  =\frac{1}{2}\left(  I+\hat
{s}\cdot \overrightarrow{\sigma}\right)  ,\\
\hat{u}^{\downarrow}\left(  \hat{s}\right)   &  =\frac{1}{2}\left(  I-\hat
{s}\cdot \overrightarrow{\sigma}\right)  ,
\end{align}
where $I$ is the unit matrix, by making use of the equality
\begin{align}
 e^{\frac{i}{\hbar^{2}}\overrightarrow{\Delta \mu}\cdot \vec{s}t}%
c_{ps^{\prime}}e^{-\frac{i}{\hbar^{2}}\overrightarrow{\Delta \mu}\cdot \vec{s}%
t}=\sum_{s^{\prime \prime}}\left[  e^{-\frac{i}{\hbar}\frac{\Delta \mu}{2}t}%
\hat{u}^{\uparrow}\left(  \hat{s}\right)  +e^{\frac{i}{\hbar}\frac{\Delta \mu
}{2}t}\hat{u}^{\downarrow}\left(  \hat{s}\right)  \right]  _{s^{\prime
}s^{\prime \prime}}c_{ps^{\prime \prime}}.
\end{align}

The leads and the island are supposed to be in thermal equilibrium, so that
$\langle c_{ks^{\prime}}^{\dagger}c_{k^{\prime}s^{\prime \prime}}\rangle
_{\circ}=f\left(  \epsilon_{ks^{\prime}}\right)  \delta_{kk^{\prime}}%
\delta_{s^{\prime}s^{\prime \prime}}$, with Fermi-Dirac distribution $f\left(
\epsilon \right)  \equiv \left(  1+e^{\beta \epsilon}\right)  ^{-1}$, where
$\beta$ is the inverse temperature. Using the expression for the matrix
elements
\begin{align}
\left[  U\left(  \theta_{\alpha}\right)  u^{s^{\prime \prime}}\left(  \hat
{s}\right)  U\left(  \theta_{\alpha}\right)  ^{\dagger}\right]  _{s^{\prime
}s^{\prime}}  &  =\frac{1}{2}\left(  1+s^{\prime}s^{\prime \prime}\hat{s}%
\cdot \overrightarrow{m}_{\alpha}\right)  ,\\
\text{with }s^{\prime},s^{\prime \prime}  &  \in \{ \uparrow,\downarrow
\}=\{+,-\},\nonumber
\end{align}
the rate of change of the number of electrons on the island reads
\begin{align}
\left \langle \frac{dn_{N}}{dt}\right \rangle _{n_{N}=m}  &  =\sum_{\alpha
s^{\prime \prime}}\frac{1}{2e^{2}}\left(  G_{\alpha}+s^{\prime \prime}P_{\alpha
}G_{\alpha}\hat{s}\cdot \overrightarrow{m}_{\alpha}\right)  \times
\label{dndt}\\
&  \left[  -F\left(  -E_{m-1}+E_{m}-\mu_{cF\alpha}+s^{\prime \prime}%
\frac{\Delta \mu}{2}\right) +F\left(  E_{m}-E_{m+1}+\mu_{cF\alpha}-s^{\prime \prime}\frac
{\Delta \mu}{2}\right)  \right],\nonumber
\end{align}
where $F\left(  \epsilon \right)  \equiv \epsilon \left(  1-e^{-\beta \epsilon
}\right)  ^{-1}$ and $E_{m}\equiv e^{2}\left(  m-C_{G}V_{G}/e\right)  ^{2}%
/2C$. The relation between the up and down spin conductances ($G_{\alpha
}^{\uparrow \uparrow}$ and $G_{\alpha}^{\downarrow \downarrow}$) and the
tunneling coefficients is $G_{\alpha}^{ss}=\left(  \pi e^{2}/\hbar \right)
\rho_{N}\rho_{F\alpha s}\left \vert T_{s}^{\alpha}\right \vert ^{2}$, where
$\left \vert T_{s}^{\alpha}\right \vert ^{2}$ is the value of $|T_{kqs}^{\alpha
}|^{2}$ at the Fermi energy averaged over all the modes. $\rho_{F\alpha s}$ is
the spin-dependent density of states in ferromagnet $\alpha$.

In the low-bias regime considered here we can linearize Eq. (\ref{dndt}) in
$\Delta \mu$ and $\mu_{cF\alpha}$. The resulting expression for the rate for
electron tunneling through contact $\alpha$, increasing the excess number of
electrons $n_{N}$ from $``0"$ to $``1"$, is denoted by $\Gamma_{\alpha
}^{0\rightarrow1}$. The analogous rate for removing one electron when $n_{N}$
is $``1"$ is $\Gamma_{\alpha}^{1\rightarrow0}$. Explicitly, we find%

\begin{align}
&  \Gamma^{0\rightarrow1}_{\alpha}(V,V_{G},\overrightarrow{\Delta \mu})=
\frac{G_{\alpha}}{e^{2}}F\left(  E_{0}-E_{1}\right)+\frac{G_{\alpha}}{e^{2}} F^{\prime}\left(  E_{0}-E_{1}\right)  \left(
-\mu_{cF\alpha}+\frac{\Delta \mu}{2}P_{\alpha} \hat{s}\cdot \overrightarrow
{m}_{\alpha}\right)  , \\
&  \Gamma^{1\rightarrow0}_{\alpha}(V,V_{G},\overrightarrow{\Delta \mu})=
\frac{G_{\alpha}}{e^{2}}F\left(  E_{1}-E_{0}\right) 
  -\frac{G_{\alpha}}{e^{2}} F^{\prime}\left(  E_{1}-E_{0}\right)  \left(
-\mu_{cF\alpha}+\frac{\Delta \mu}{2}P_{\alpha} \hat{s}\cdot \overrightarrow
{m}_{\alpha}\right)  .
\end{align}

Now that we have determined the tunneling rates we can write down the master
equation for electron transport in the orthodox model. We consider a regime in
which $eV\ll k_{B}T\ll e^{2}/2C$, and restrict ourselves to a gate voltage
range for which the excess number of electrons $n_{N}$ alternates between
$``0"$ and $``1"(0<C_{G}V_{G}<e),$ knowing that the results will periodically
repeat with this period. The center of the Coulomb oscillation for transitions
between $n_{N}=``0"$ and $``1"$ electrons is at $C_{G}V_{G}=e/2$.

The steady state on the island is characterized by a constant spin
accumulation (to be determined below) and the probabilities $p_{0}$ and
$p_{1}$ that there are $``0"$ or $``1"$ excess electrons. We have $p_{0}%
+p_{1}=1$. The rate equation for the probabilities is
\begin{align}
dp_{n}/dt    =-p_{n}\left(  \Gamma^{n\rightarrow n+1}+\Gamma^{n\rightarrow
n-1}\right)+p_{n+1}\Gamma^{n+1\rightarrow}+p_{n-1}\Gamma^{n-1\rightarrow n}.
\end{align}
From the condition of detailed balance, $p_{0}\Gamma^{0\rightarrow1}%
=p_{1}\Gamma^{1\rightarrow0}$, we find
\begin{align}
\label{pzero}   p_{0}(V,V_{G},\overrightarrow{\Delta \mu})=f\left(
E_{0}-E_{1}\right)  +\frac{\beta f\left(  E_{0}-E_{1}\right)  f\left(
E_{1}-E_{0}\right)  }{G_{1}+G_{2}}\sum_{\alpha}\left(  G_{\alpha}\mu_{cF\alpha}-P_{\alpha}G_{\alpha}%
\frac{\Delta \mu}{2}\hat{s}\cdot \overrightarrow{m}_{\alpha}\right).
\end{align}
The expression for the conductance of the SV-SET as a function of the spin
accumulation can now be calculated and reads
\begin{align}
G(V,V_{G},\overrightarrow{\Delta \mu})  &  =-ep_{0}\Gamma_{1}^{0\rightarrow
1}+ep_{1}\Gamma_{1}^{1\rightarrow0}\nonumber \\
&  =\frac{G_{1}G_{2}}{G_{1}+G_{2}}\frac{\beta \left(  E_{0}-E_{1}\right)
}{2\sinh \beta \left(  E_{0}-E_{1}\right)  }\left[  1-\frac{\Delta \mu}{2eV}\hat{s}\cdot \left(  P_{1}\overrightarrow
{m}_{1}-P_{2}\overrightarrow{m}_{2}\right)  \right].  \label{cond}%
\end{align}

\subsection{Spin accumulation}

\label{spinacc} The steady-state spin accumulation is found by setting the
total rate of change of $\vec{s}$ to zero. There are several contributions to
the dynamics of the spin accumulation:
\begin{align}
\left \langle \frac{d\vec{s}}{dt}\right \rangle  &  =p_{0}\left \langle
\frac{d\vec{s}}{dt}\right \rangle _{n_{N}=0}+p_{1}\left \langle \frac{d\vec{s}%
}{dt}\right \rangle _{n_{N}=1}\nonumber \\
&  +\sum_{\alpha}\left \langle \  \frac{d\vec{s}}{dt}\right \rangle
_{\mathrm{ex}\alpha}+\left \langle \  \frac{d\vec{s}}{dt}\right \rangle
_{\mathrm{magn}}+\left \langle \  \frac{d\vec{s}}{dt}\right \rangle
_{\mathrm{sf}}. \label{dsdt}%
\end{align}
The first two terms are due to the tunneling processes, the remaining ones to
exchange, external magnetic fields and spin flip. We start from
\begin{equation}
\frac{d\vec{s}}{dt}=\frac{i}{\hbar}\left[  H_{T},\vec{s}\right]  ,
\end{equation}
with an expectation value that to second order in $H_{T}$ reads
\[
\left \langle \frac{d\vec{s}\left(  t\right)  }{dt}\right \rangle =\frac
{i}{\hbar}\int \limits_{-\infty}^{t}dt^{\prime}\left \langle \left[  \frac
{d\vec{s}\left(  t\right)  }{dt},H_{T}\left(  t^{\prime}\right)  \right]
\right \rangle .
\]
The spin current (rate of change of the spin angular momentum) due to
tunneling when $m$ excess electrons are on the island reads (\textit{cf}. Eq.
(\ref{dndt}))
\begin{align}
\label{spincurr} &  \left \langle \  \frac{d\vec{s}}{dt}\right \rangle _{n_{N}%
=m}=\frac{\hbar}{4e^{2}}\sum_{\alpha s^{\prime \prime}}\left(  G_{\alpha
}s^{\prime \prime}\hat{s}+P_{\alpha}G_{\alpha}\overrightarrow{m}_{\alpha
}\right)  \times \nonumber \\
&  \left[  -F\left(  -E_{m-1}+E_{m}-\mu_{cF\alpha}+s^{\prime \prime}%
\frac{\Delta \mu}{2}\right)+F\left(  -E_{m+1}+E_{m}+\mu_{cF\alpha}-s^{\prime \prime}%
\frac{\Delta \mu}{2}\right)  \right] \nonumber \\
&  +\frac{\hbar}{4\pi e^{2}}\sum_{\alpha s^{\prime \prime}}P_{\alpha}G_{\alpha
}s^{\prime \prime}\left(  \overrightarrow{m}_{\alpha}\times \hat{s}\right)
\times \nonumber \\
&  \left(  \int d\epsilon_{1}\int^{\prime}d\epsilon_{2}\frac{f\left(
\epsilon_{1}\right)  \left(  1-f\left(  \epsilon_{2}\right)  \right)
}{\left(  \epsilon_{2}-\epsilon_{1}+E_{m-1}-E_{m}+\mu_{cF\alpha}%
-s^{\prime \prime}\frac{\Delta \mu}{2}\right)  }\right. \nonumber \\
&  \left.  -\int d\epsilon_{1}\int^{\prime}d\epsilon_{2}\frac{f\left(
\epsilon_{2}\right)  \left(  1-f\left(  \epsilon_{1}\right)  \right)
}{\left(  \epsilon_{2}-\epsilon_{1}+E_{m}-E_{m+1}+\mu_{cF\alpha}%
-s^{\prime \prime}\frac{\Delta \mu}{2}\right)  }\right)  ,
\end{align}
where the prime denotes a principal value integral. Here we used the
relation:
\begin{align}
\left[  U\left(  \theta_{\alpha}\right)  u^{s^{\prime \prime}}\left(  \hat
{s}\right)  \sigma^{i}U\left(  \theta_{\alpha}\right)  ^{\dagger}\right]
_{s^{\prime}s^{\prime}}    =\frac{1}{2}s^{\prime \prime}\hat{s}
+\frac{1}{2}s^{\prime}\overrightarrow{m}_{\alpha}+\frac{1}{2}is^{\prime
}s^{\prime \prime}\left(  \overrightarrow{m}_{\alpha}\times \hat{s}\right)  ,\\
\text{with }s^{\prime},s^{\prime \prime}  &  \in \{ \uparrow,\downarrow
\}=\{+,-\}.\nonumber
\end{align}

To first order in the small induced energy shifts the exchange Hamiltonian
$H_{\mathrm{ex}\alpha}$ modifies the unpaired spins as
\begin{equation}
\left.  \frac{d\vec{s}}{dt}\right \vert _{\mathrm{ex}\alpha}=\frac{i}{\hbar
}\left[  H_{\mathrm{ex}\alpha},\vec{s}\left(  t\right)  \right]  ,
\end{equation}
which results in a precession:
\begin{equation}
\left \langle \frac{d\vec{s}}{dt}\right \rangle =\frac{1}{\hbar M}\sum
_{m}\left(  \Delta \epsilon_{\alpha m\uparrow}-\Delta \epsilon_{\alpha
m\downarrow}\right)  \left \langle \vec{s}\right \rangle \times \vec{m}_{\alpha},
\label{precession}%
\end{equation}
where $M$ is the number of transport channels in the normal metal and the
energy shifts $\Delta \epsilon$ are found in Eq. (\ref{shifts}).

The conductance parameters of an F$|$N contact are\cite{Brataas}
\begin{equation}
G_{\alpha}^{ss^{\prime}}\equiv \frac{e^{2}}{h}\sum \limits_{nm}\left(
\delta_{nm}-r_{s\alpha}^{nm}\left(  r_{s^{\prime}\alpha}^{nm}\right)  ^{\ast
}\right)  ,(s,s^{\prime}\in \uparrow,\downarrow).\label{gmix}%
\end{equation}
Here $n$ and $m$ denote the transport channels in the normal metal and
$r_{\uparrow \alpha}^{nm}$ and $r_{\downarrow \alpha}^{nm}$ are the
corresponding spin-dependent reflection coefficients. The contact conductances
for spin-up and spin-down electrons are $G_{\alpha}^{\uparrow \uparrow \text{ }%
}$ and $G_{\alpha}^{\downarrow \downarrow}$ and the mixing conductance
$G_{\alpha}^{\uparrow \downarrow}$ governs the transverse spin currents that
are absorbed and reflected by the ferromagnet $\alpha$. The current polarized
normal to the magnetization but in the plane of  $\vec{s}$ and $\vec
{m}_{\alpha}$ is proportional to $\operatorname{Re}G_{\alpha}^{\uparrow
\downarrow}$ and describes the spin-transfer to the magnet, thereby
dissipating the spin accumulation. In the case of tunnel junctions
$\operatorname{Re}G_{\alpha}^{\uparrow \downarrow}\rightarrow G_{\alpha}/2$.
The out-of-the $\vec{s}$,$\vec{m}_{\alpha}$ plane spin current is caused by
reflection processes that make spins precess around $\vec{m}_{\alpha}$ and is
proportional to $\operatorname{Im}G^{\uparrow \downarrow}$. This mixing
conductance has been evaluated from first principles for various contact
materials and is small for intermetallic interfaces because positive and
negative contributions in the space spanned by the transport channels average
out.\cite{Xia} However, there is no general reason that $\operatorname{Im}%
G^{\uparrow \downarrow}$ should be smaller than $G$ or $\operatorname{Re}%
G^{\uparrow \downarrow}.$ It is known to be quite large for the Fe$|$InAs
interface\cite{Zwierzycki} and found to be very significant for the
magnetization dynamics of MgO magnetic tunnel junctions.\cite{Tulapurkar} For
a simple model barrier discussed in Appendix \ref{imgmodel}, we find the value
$\operatorname{Im}G^{\uparrow \downarrow}/G=-0.26$. Using the relation between
the reflection phases and the energy shifts as derived in Eq. (\ref{shifts}),
we can rewrite the contribution given in Eq. (\ref{precession}) in terms of
the imaginary part of the mixing conductance as (cf. Ref.
\onlinecite{Wetzels})
\begin{equation}
\left \langle \frac{d\vec{s}}{dt}\right \rangle _{\mathrm{ex}\alpha}%
=\frac{\operatorname{Im}G_{\alpha}^{\uparrow \downarrow}}{\rho_{N}e^{2}%
}\overrightarrow{m}_{\alpha}\times \left \langle \vec{s}\right \rangle .
\end{equation}

The spin accumulation can also be affected by a magnetic field $\vec{B}$,
which can either be externally applied, a stray field from the ferromagnets,
or an internal anisotropy field. The spin accumulation induced by this
magnetic field can safely be neglected, but the induced precession of the spin
accumulation is relevant, and is given by
\begin{equation}
\left \langle \frac{d\vec{s}}{dt}\right \rangle _{\mathrm{magn}}=\frac{g\mu_{B}%
}{\hbar}\vec{B}\times \left \langle \vec{s}\right \rangle .
\end{equation}

Finally, spin-flip relaxation in the normal metal is taken into account by
spin-accumulation decay with a spin-flip relaxation time $\tau_{\mathrm{sf}%
},$
\begin{equation}
\left \langle \frac{d\vec{s}}{dt}\right \rangle _{\mathrm{sf}}=-\frac
{\left \langle \vec{s}\right \rangle }{\tau_{\mathrm{sf}}}.
\end{equation}

Combining the terms in Eq. (\ref{dsdt}), the spin accumulation should fulfil
the stationary state condition:%
\begin{align}
\left \langle \frac{d\vec{s}}{dt}(V,V_{G})\right \rangle  &  =\frac{\hbar
}{2e^{2}}\frac{\beta \left(  E_{0}-E_{1}\right)  }{2\sinh \beta \left(
E_{0}-E_{1}\right)  }\times \nonumber \\
&  \left[  \frac{G_{1}G_{2}}{G_{1}+G_{2}}eV\left(  P_{1}\vec{m}_{1}-P_{2}%
\vec{m}_{2}\right)-\left(  G_{1}+G_{2}\right)  \frac{\Delta \mu}{2}\left(  \hat
{s}+\left(  \hat{s}\cdot \vec{b}\right)  \vec{b}\right)  \right] \nonumber \\
&  +\frac{g\mu_{B}}{\hbar}\vec{B}_{\mathrm{eff}}\times \left \langle \vec
{s}\right \rangle -\frac{\left \langle \vec{s}\right \rangle }{\tau_{\mathrm{sf}%
}}=0, \label{squasieq}%
\end{align}
where
\begin{align}
\vec{b}\equiv \frac{P_{1}G_{1}}{G_{1}+G_{2}}\overrightarrow{m}_{1}+\frac
{P_{2}G_{2}}{G_{1}+G_{2}}\overrightarrow{m}_{2}.
\end{align}
The total effective magnetic field $\vec{B}_{\mathrm{eff}}$ consists of the
external magnetic field and contributions from the exchange effects X1 and X2,
and reads%
\begin{equation}
\vec{B}_{\mathrm{eff}}(V_{G})=\vec{B}+\vec{B}_{X1}+\vec{B}_{X2}\left(
V_{G}\right),
\end{equation}
\begin{align}
\text{with }\vec{B}_{X1}  &  =\frac{\hbar}{\rho_{N}g\mu_{B}e^{2}}\sum_{\alpha
}\operatorname{Im}G_{\alpha}^{\uparrow \downarrow}\overrightarrow{m}_{\alpha},\\
\vec{B}_{X2}\left(  V_{G}\right)   &  =-\frac{\hbar}{2\rho_{N}g\mu_{B}e^{2}%
}\left(  G_{1}+G_{2}\right)  \vec{b}\label{BX2}\\
&  \left[  \frac{1}{\pi}f\left(  E_{0}-E_{1}\right)  \int d\epsilon f^{\prime
}\left(  \epsilon \right)  \eta \left(  \epsilon+E_{0}-E_{-1},\frac{e^{2}}%
{C}\right)  \right. \nonumber \\
&  +\frac{1}{\pi}f\left(  E_{1}-E_{0}\right)  \left.  \int d\epsilon
f^{\prime}\left(  \epsilon \right)  \eta \left(  \epsilon+E_{1}-E_{0}%
,\frac{e^{2}}{C}\right)  \right]. \nonumber
\end{align}
Here we introduced\cite{Koenig}
\begin{align}
\label{defeta}\eta \left(  \varepsilon,U\right)   &  \equiv \int^{\prime}%
d\varpi \left(  \frac{1-f\left(  \varpi \right)  }{\varpi-\varepsilon}%
+\frac{f\left(  \varpi \right)  }{\varpi-\varepsilon-U}\right) \\
&  =-\operatorname{Re}\left[  \Psi \left(  \frac{1}{2}+\frac{i\beta \varepsilon
}{2\pi}\right)  -\Psi \left(  \frac{1}{2}+\frac{i\beta \left(  \varepsilon
+U\right)  }{2\pi}\right)  \right]  ,\nonumber
\end{align}
where $\Psi \left(  z\right)  $ is the Digamma function. In appendix
\ref{virttun} we discuss the derivation of the expression for $\vec{B}_{X2}$
in more detail and comment on the differences compared to the case of a
single-level quantum dot.

\section{Results and discussion}

\label{results}

\begin{figure}[tbh]
\centering
\includegraphics[width=3.375in]{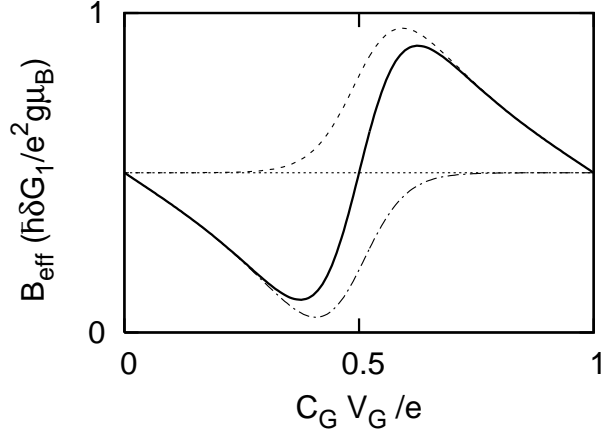}\caption{ The effective
magnetic field strength $\left \vert \vec{B}_{\mathrm{eff}}\right \vert $ as a
function of gate voltage (solid line) for a spin valve in the parallel
configuration. The parameters are $G_{1}=G_{2},$ $P_{1}=P_{2}=0.7$,
$\operatorname{Im}G_{1}^{\uparrow \downarrow}=\operatorname{Im}G_{2}%
^{\uparrow \downarrow}=G_{1}/4$ and $e^{2}/\left(  2C\right)  =10k_{B}T$. The
imaginary part of the mixing conductance gives a constant offset (dotted). The
dot-dashed/dashed curves show the effective field for zero and one excess
electron on the island. }%
\label{exch}%
\end{figure}

Fig. \ref{exch} shows the magnitude of the total effective magnetic field
$\vec{B}_{\mathrm{eff}}$ as a function of gate voltage (solid line) for a
symmetric spin valve with parallel magnetizations (it vanishes for the
antiparallel configuration) and a polarization $P_{1}=0.7$. The X1 term is a
constant that does not depend on gate voltage (dotted line). $\vec
{B}_{\mathrm{X2}}$ vanishes when $C_{G}V_{G}$ equals $0,e/2$ and $e$. At these
points, contributions from incoming and outgoing electrons cancel each other
(see appendix \ref{virttun}). The curve repeats as a function of gate voltage
with period $e/C_{G}$. The spin accumulation on the island found from Eq. (\ref{squasieq}%
) tends to suppress the current through the system. Spin-flip and exchange
effects that dissipate or dephase the spin-accumulation therefore increase the
conductance. As a reference we list here the conductance $G\left(
\theta \right)  $ for a spin valve \textit{without interaction}, with equal
conductance parameters for the left and the right tunneling barrier
$G_{1}=G_{2},$ $P_{1}=P_{2}$:
\begin{equation}
G\left(  \theta \right)  =\frac{G_{1}}{2}\left(  1-\frac{P_{1}^{2}G_{1}\left(
G_{1}+2G_{sf}\right)  \sin^{2}\theta/2}{G_{1}+2G_{sf}+\left(  2\cos
\frac{\theta}{2}\operatorname{Im}G_{1}^{\uparrow \downarrow}\right)  ^{2}%
}\right)  .\label{spin-valve conductance}%
\end{equation}
The final result for the symmetric spin-valve \textit{with interaction} can be
obtained simply from this expression by the substitutions:
\begin{align}
G_{1} &  \rightarrow \frac{\beta \left(  E_{0}-E_{1}\right)  }{2e\sinh
\beta \left(  E_{0}-E_{1}\right)  }G_{1},\\
\operatorname{Im}G_{1}^{\uparrow \downarrow} &  \rightarrow \frac{e^{2}}{\hbar
}\frac{\rho_{N}g\mu_{B}B_{\mathrm{eff}}}{2\cos \frac{\theta}{2}}.
\end{align}
For nonmagnetic contacts ($P_{1}=0$) this result reduces to the known
expression for normal metal single-electron transistors.\cite{Kulik}

\begin{figure}[tbh]
\centering
\includegraphics[width=3.375in]{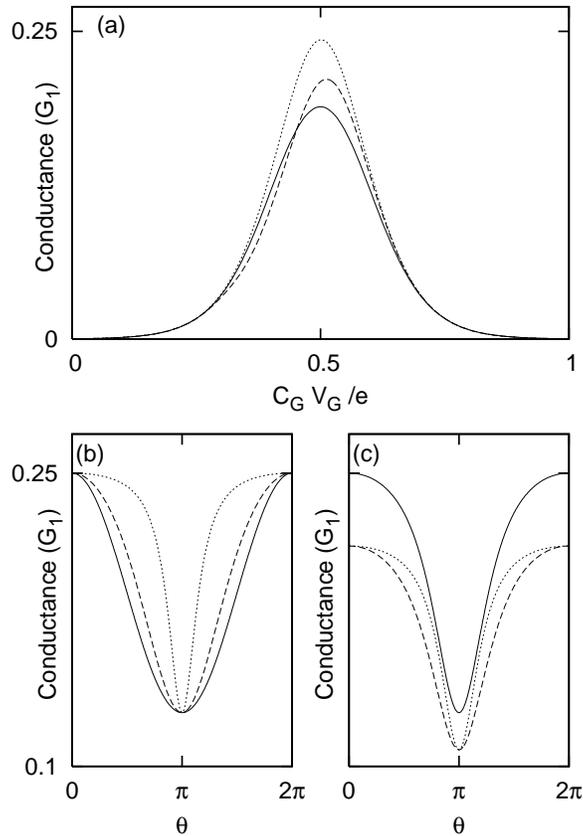}\caption{(a) Coulomb
oscillations at fixed angle $\theta=\pi/2$ for a symmetric SV-SET with ratio
$\operatorname{Im}G_{1}^{\uparrow \downarrow}/G_{1}=0$ (solid), $0.25$ (dashed)
and $1$ (dotted) in units of $G_{1}$. The polarization $P$ is $0.7$. (b)
Conductance as a function of the angle for the same parameters as in a), with
$C_{G}V_{G}$ fixed to 0.5. (c) Conductance as a function of $\theta$ with
$\operatorname{Im}G_{1}^{\uparrow \downarrow}/G_{1}=0.25$. Results are shown
for $C_{G}V_{G}$ equal to $0.45$ (dashed)$,0.5$ (solid) and $0.55$ (dotted).}%
\label{Cosc}%
\end{figure}As shown in Fig. \ref{Cosc}(a), changing the relative strengths of
X1 and X2, or, since the X2 contribution is proportional to the polarization
of the leads, $\operatorname{Im}G_{\alpha}^{\uparrow \downarrow}/P_{\alpha
}G_{\alpha},$ qualitatively modifies the current profile of the Coulomb
oscillations. The constant offset given by $B_{X1}$ skews the exchange field
around $C_{G}V_{G}=e/2$, causing asymmetric conductance curves. When the
offset starts to dominate the symmetry gets restored. The X2 contribution
vanishes when the Coulomb blockade is lifted $\left(  C_{G}V_{G}=e/2\right)
$, so the angular dependence of the conductance for different values of
$\operatorname{Im}G_{1}^{\uparrow \downarrow}/G_{1}$ in Fig. \ref{Cosc}(b)
reflects only the X1 effect. The curve is a simple cosine for
$\operatorname{Im}G_{1}^{\uparrow \downarrow}=0$, but is sharpened for larger
$\operatorname{Im}G_{1}^{\uparrow \downarrow}$ because of the dephasing of the
spin accumulation occurring for noncollinear angles. In Fig. \ref{Cosc}(c)
$\operatorname{Im}G_{1}^{\uparrow \downarrow}/G_{1}$ is fixed to $0.25$ and
curves are plotted for different values of the gate voltage. It can be seen
that the angular dependence differs because the X2 depends on $V_{G}$ in an
asymmetric way around $C_{G}V_{G}=e/2$.

\begin{figure}[tbh]
\centering
\includegraphics[width=3.375in]{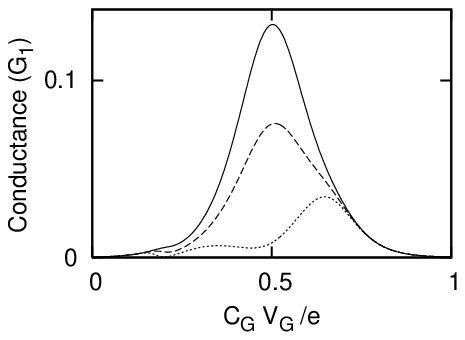}\caption{Conductance as a function of 
gate voltage for a symmetric SV-SET with $\theta=0.9\pi$ and $\operatorname{Im}G_{1}^{\uparrow \downarrow}/G_{1}=0.15$. Curves are shown for $P_{1}=0.7$ (solid), $P_{1}=0.85$ (dashed) $P_{1}  =1$ (dotted).}
\label{highP}%
\end{figure}

As can be seen in Fig. \ref{highP}, the shape of the Coulomb oscillation can 
develop minima when the polarization is high and the magnetizations are nearly antiparallel. 
At the values of gate voltage where the X1 and X2 exchange effects cancel, the spin 
accumulation is not dephased and the conductance is suppressed. 

Fig. \ref{mu} shows results for the conductance and spin accumulation as a
function of applied magnetic field in the $x$ (solid line), $y$ (dashed) and
$z$ (dotted) directions. The spin valve is again symmetric with $P_{1}=0.7$
and $\operatorname{Im}G_{1}^{\uparrow \downarrow}=G_{1}/4$. The angle $\theta$
is fixed to $\pi/2$ and $C_{G}V_{G}=e/2$. The conductance then depends only on
the $x$-component of the spin accumulation (see Eq. (\ref{cond})). Without
applied magnetic field, the spin accumulation has components in the $x$ and
$y$ directions, while $\vec{B}_{\mathrm{eff}}$ is in the $y$-direction. The
results can be understood in terms of the dephasing of the spin accumulation
by the magnetic-field induced precession that, for sufficiently large and
non-collinear magnetic fields, quenches the spin accumulation. This
\textquotedblleft Hanle\textquotedblright \ effect is responsible for the
conductance minimum at negative applied magnetic field in the $y$ direction.
In \ref{mu}(c) only two curves are visible because the curves for magnetic
fields in the $x$ and $y$ direction overlap.\begin{figure}[tbh]
\centering
\includegraphics[width=3.375in]{sv2_F_mu.eps}\caption{(a) Conductance as a
function of a magnetic field applied along the the $x$ (solid), $y$ (dashed)
or $z$ (dotted) direction in units of $G_{1}$. The SV-SET has symmetric
junction parameters, with polarizations $P_{1}=0.7$ and $\operatorname{Im}%
G_{1}^{\uparrow \downarrow}=G_{1}/4$. The magnetizations are fixed to
$\protect\overrightarrow{m}_{1/2}=(\pm1,0,1)/\sqrt{2}$, yielding an angle $\theta
=\pi/2$ and $C_{G}V_{G}=e/2$. (b,c,d) The $x$, $y$ and $z$ components of the
spin accumulation for the same parameters. The curves in (c) for magnetic
fields in the $x$ and $y$ directions overlap.}%
\label{mu}%
\end{figure}

\section{Summary}

We studied the transport properties of single-electron spin valve transistors
as a function of the magnetization configurations in the orthodox model of the
Coulomb blockade. Two types of exchange effects between the spin accumulation
on the island and the lead magnetizations play a role: a nonlocal interface
exchange effect (X1) and exchange due to virtual tunneling processes (X2). For
metallic dots these two effects are found to be of comparable magnitude. We
predict that a line shape analysis of the Coulomb oscillation peaks should
help to experimentally disentangle the two contributions. Additional
information can be obtained by the Hanle effect.

\section{Acknowledgements}

We would like to thank J. K\"{o}nig, J. Martinek, M. Braun and Y.V. Nazarov
for useful discussions. This research was supported by the NWO and by the DFG
via the SFB 689, and supported in part by the National Science Foundation
under Grant No. PHY99-07949.

\appendix

\section{Energy shifts}

\label{appshifts}

\begin{figure}[tbh]
\centering
\includegraphics[width=1.375in]{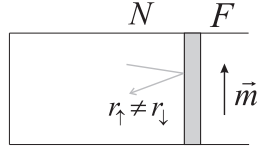}\caption{A normal metal
island with tunnel contact to a ferromagnetic lead.}%
\label{F_shifts}%
\end{figure}

Let us consider a normal metal island in contact to a ferromagnet by a tunnel
barrier (See Fig. \ref{F_shifts}) without Coulomb interaction. We wish to
calculate the spin-dependent shifts of the energy levels due to the presence
of the F$|$N contact. In Ref. \onlinecite{Cottet} an analogous calculation was
done for a ballistic one-dimensional quantum wire. Here we consider an island
in the quasi-classical regime, \textit{i.e.} the de Broglie wavelength is much
smaller than the size of the island.

The Bohr-Sommerfeld quantization rule \cite{Landau}
\begin{equation}
\frac{1}{\hbar}\oint p^{m}\left(  x\right)  dx+\phi_{0}^{m}+\phi_{s}^{m}%
=2\pi \left(  n+\frac{1}{2}\right)  \label{Bohr-Sommerfeld}%
\end{equation}
can be used to find the energy shifts, where $p^{m}\left(  x\right)  $ is the
classical momentum for an electron in mode $m$, and $n$ is an integer. The
integral is over a whole period of the classical motion in the quasi-classical
region. The total phase shift due to the reflections at the turning points is
$\phi_{0}^{m}+\phi_{s}^{m}$, where $\phi_{0}^{m}$ is the spin-independent
phase shift picked up during the reflections from the boundaries for an
isolated island without contact to the ferromagnet. The small spin-dependent
phase shift $\phi_{s}^{m}\ll1$ arises from the weak coupling to the
ferromagnet. The phase shifts have to be computed quantum mechanically via the
spin-dependent reflection coefficients $r_{s}^{mm}$ for mode $m$ at an
interface that is assumed to be specular (see also Appendix \ref{imgmodel}%
).

From Eq. (\ref{Bohr-Sommerfeld}), we see that increasing the quantum number
$n$ by one corresponds to introducing an extra phase period that increases the
kinetic energy by $M/\rho_{N},$ where $\rho_{N}$ is the density of states of
the island and $M$ is the number of modes. The energy shift for an electron in
mode $m$ is therefore, to linear order in $\phi_{s}^{m}$,
\begin{equation}
\Delta \epsilon_{ms}=-\frac{M}{\rho_{N}}\frac{\phi_{s}^{m}}{2\pi},
\label{shifts}%
\end{equation}
The effect of the interface on the island states can be taken into account by
introducing an effective Hamiltonian as in Eq. (\ref{Hex}). In the case of a
spin-independent tunneling barrier to a ferromagnet, the spin-splitting of the
energy levels is small, of the same order as the transmission probability (see
App. \ref{imgmodel}).

\section{Rectangular barriers}

\label{imgmodel} Here we evaluate the spin-mixing conductance $G^{\uparrow
\downarrow}$ for a model barrier, giving more details of the results of Ref.
\onlinecite{Wetzels}. We consider a smooth rectangular barrier between a
normal metal and a Stoner-model ferromagnet. The solution of the
Schr\"{o}dinger equation for spin $s$ in the normal metal, $\psi_{s}%
^{m}\left(  x,y,z\right)  $ can be used to determine the reflection
coefficients $r_{s}^{mm}$ for each mode $m$. It reads
\[
\psi_{s}^{m}\left(  x,y,z\right)  =\frac{\chi^{m}\left(  x,y,z\right)  }%
{\sqrt{k_{N}^{m}}}\left(  e^{ik_{N}^{m}x}+r_{s}^{mm}e^{-ik_{N}^{m}x}\right)
,
\]
where $\chi^{m}\left(  y,z\right)  $ is the transverse wave function and
$k_{N}^{m}$ is the longitudinal wave number for mode $m$ in the normal metal.
In terms of the wave numbers in the normal metal $k_{N}^{m}$, barrier
$k_{B}^{m}$ and ferromagnet $k_{Fs}^{m}$ for a given energy, the reflection
coefficient for mode $m$ at the barrier reads
\begin{align}
r_{s}^{mm}=\rho(k_{N}^{m},k_{B}^{m}) +e^{2iak_{B}^{m}}\tau(k_{N}^{m},k_{B}^{m})\rho(k_{B}^{m},k_{Fs}^{m}%
)\tau(k_{B}^{m},k_{N}^{m}),
\end{align}
where $a$ is the barrier thickness and
\begin{align}
\tau \left(  k_{1},k_{2}\right)   &  \equiv \frac{2\sqrt{k_{1}k_{2}}}%
{k_{1}+k_{2}},\\
\rho \left(  k_{1},k_{2}\right)   &  \equiv \frac{k_{1}-k_{2}}{k_{1}+k_{2}}.
\end{align}
For a tunneling barrier, $k_{B}^{m}$ is imaginary and the spin dependent
correction to the reflection coefficient is exponentially small in the barrier thickness.

For a numerical estimate we use a Fermi energy in the normal metal of 2.6 eV,
a barrier height of 3 eV and barrier thickness of $a=1
\operatorname{nm}%
$. The Fermi momenta in the ferromagnet are taken to be $k_{F\uparrow
}=1.09\mathring{A}^{-1}$ and $k_{F\downarrow}=0.42\mathring{A}^{-1}$
(characteristic for Fe, see Ref. \onlinecite{Slonczewski}).

For the spin-mixing conductance $G^{\uparrow \downarrow}$, Eq. (\ref{gmix}), we
find that $\operatorname{Im}G^{\uparrow \downarrow}/G=-0.26$ for this choice of
parameters. The effective field due the interface exchange effect is therefore
not negligible compared to the conductance parameters. More realistic
electronic structure calculations\cite{Zwierzycki} should be carried out to
obtain better estimates.

\section{X2 exchange in classical dots }
\label{virttun} Here we present more details concerning the derivation of Eq.
(\ref{BX2}) for the effective exchange field X2 in classical SV-SET's, that
complement the derivation in Refs. \onlinecite{Koenig} for single-level
quantum dots. Since the model is periodic in the gate voltage with period
$e/C_{G}$, we restrict our discussion to the range $0<C_{G}V_{G}<e$. From Eq.
(\ref{dsdt}), the contributions from virtual tunneling processes to the rate
of change of $\vec{s}$ then read:
\[
\left \langle \frac{d\vec{s}}{dt}\right \rangle _{X2}=p_{0}\left \langle
\frac{d\vec{s}}{dt}\right \rangle _{X2,n_{N}=0}+p_{1}\left \langle \frac
{d\vec{s}}{dt}\right \rangle _{X2,n_{N}=1}.
\]

Using the spin currents from Eq. (\ref{spincurr}), we obtain, \textit{{e.g.}%
}:
\begin{align}
\left \langle \frac{d\vec{s}}{dt}\right \rangle _{X2,n_{N}=0}   &=\frac{\hbar
}{4\pi e^{2}}\sum_{\alpha s^{\prime \prime}}P_{\alpha}G_{\alpha}s^{\prime
\prime}\left(  \overrightarrow{m}_{\alpha}\times \hat{s}\right)  \times
\nonumber \label{dsdtX2}\\
 & \left(  \int d\epsilon_{1}\int^{\prime}d\epsilon_{2}\frac{f\left(
\epsilon_{1}\right)  \left(  1-f\left(  \epsilon_{2}\right)  \right)
}{\left(  \epsilon_{2}-\epsilon_{1}+E_{-1}-E_{0}+\mu_{cF\alpha}-s^{\prime
\prime}\frac{\Delta \mu}{2}\right)  }\right.  \nonumber \\
 & \left.  -\int d\epsilon_{1}\int^{\prime}d\epsilon_{2}\frac{f\left(
\epsilon_{2}\right)  \left(  1-f\left(  \epsilon_{1}\right)  \right)
}{\left(  \epsilon_{2}-\epsilon_{1}+E_{0}-E_{1}+\mu_{cF\alpha}-s^{\prime
\prime}\frac{\Delta \mu}{2}\right)  }\right)  .
\end{align}
The first term in brackets describes virtual processes in which an electron
tunnels out of the island, and the second term corresponds to incoming
electrons. The expressions for the energy differences are given by
\begin{align}
E_{-1}-E_{0} &  =\left(  C_{G}V_{G}+e/2\right)  e/C,\\
E_{0}-E_{1} &  =\left(  C_{G}V_{G}-e/2\right)  e/C.
\end{align}
Because of the periodicity in the gate voltage
\[
\left \langle \frac{d\vec{s}}{dt}\right \rangle _{X2,n_{N}=1}=\left \langle
\frac{d\vec{s}}{dt}\right \rangle _{X2,n_{N}=0}\text{with }V_{G}\rightarrow
V_{G}-e/C_{G}.
\]
We can now rewrite Eq. (\ref{dsdtX2}) in terms of the function $\eta \left(
\epsilon,U\right)  $, defined in Eq. (\ref{defeta}). The probabilities $p_{0}$
and $p_{1}$ are taken from Eq. (\ref{pzero}). After linearization in $V$ and
$\Delta \mu$, we arrive at the expression Eq. (\ref{BX2}).

We note the differences with the results for single-level quantum
dots;\cite{Koenig} our expression includes an additional integral over the
island states. For a single-level quantum dot, X2 is  active only when exactly
one electron resides on the dot, since there is no unpaired spin in an empty
or doubly occupied dot. In contrast, a net spin accumulation can reside on our
classical dot for all numbers of electrons. The effective magnetic field is a
sum weighted with the probabilities for $``0$\textquotedblright \ and
$``1$\textquotedblright \ electrons on the dot, which
leads to a partial cancellation of the contributions for different $n_{N}$, as
is shown in Fig. \ref{exch}.


\begin{thebibliography}{99}                                                                                               %


\bibitem {Ono}K. Ono, H. Shimada, and Y. Ootuka, J. Phys. Soc. Jpn
\textbf{66}, 1261 (1997).

\bibitem {Schelp}L. F. Schelp, A. Fert, F. Fettar, P. Holody, S. F. Lee, J. L.
Maurice, F. Petroff, and A. Vaur\`{e}s, Phys. Rev. B \textbf{56}, R5747 (1997).

\bibitem {Ralph}M. M. Deshmukh and D. C. Ralph, Phys. Rev. Lett. \textbf{89},
266803 (2002).

\bibitem {Pasupathy}A. N. Pasupathy, R. C. Bialczak, J. Martinek, J. E. Grose,
L. A. K. Donev, P. L. McEuen, and D. C. Ralph, Science \textbf{306}, 86 (2004).

\bibitem {Philip}J. Philip, D. Wang, M. Muenzenberg, P. LeClair, B. Diouf, J.
S. Moodera, and J. G. Lu, J. Magn. Magn. Mater., \textbf{272-276}, 1949 (2004).

\bibitem {Zhang}L. Y. Zhang, C. Y. Wang, Y. G. Wei, X. Y. Liu, and D.
Davidovi\'{c}, Phys. Rev. B, \textbf{72}, 155445 (2005).

\bibitem {Sahoo}S. Sahoo, T. Kontos, J. Furer, C. Hoffmann, M. Gr\"{a}ber, A.
Cottet, and C. Sch\"{o}nenberger, Nature Physics \textbf{1}, 99 (2005).

\bibitem {Yakushiji}K. Yakushiji, F. Ernult, H. Imamura, K. Yamane, S. Mitani,
K. Takanashi, S. Takahashi, S. Maekawa, and H. Fujimori, Nature Mat.
\textbf{4}, 57 (2005).

\bibitem {Bernand-Mantel}A. Bernand-Mantel, P.\ Seneor, N. Lidgi, M.
Mu\~{n}oz, V. Cros, S. Fusil, K. Bouzehouane, C. Deranlot, A. Vaures, F.
Petroff, and A. Fert, cond-mat/0601439.

\bibitem {Barnas}J. Barna\'{s} and A. Fert, Phys. Rev. Lett. \textbf{80}, 1058 (1998).

\bibitem {Majumdar}K. Majumdar and S. Hershfield, Phys. Rev. B \textbf{57},
11521 (1998).

\bibitem {Korotkov}A. N. Korotkov and V. I. Safarov, Phys. Rev. B \textbf{59},
89 (1999).

\bibitem {Brataasnib}A. Brataas, Yu. V. Nazarov, J. Inoue, and G. E. W. Bauer,
Eur. Phys. J. B \textbf{9}, 421 (1999).

\bibitem {Brataas and Wang}A. Brataas and X. H. Wang, Phys. Rev. B
\textbf{64}, 104434 (2001).

\bibitem {Balents}L. Balents and R. Egger, Phys. Rev. B \textbf{64}, 035310 (2001).

\bibitem {Bena}C. Bena, and L. Balents, Phys. Rev. B \textbf{65}, 115108 (2002).

\bibitem {Koenig}J. K\"{o}nig and J. Martinek, Phys. Rev. Lett. \textbf{90},
166602 (2003); M. Braun, J. K\"{o}nig, and J. Martinek, Phys. Rev. B
\textbf{70}, 195345 (2004); J. K\"{o}nig, J. Martinek, J. Barna\'{s}, and G.
Sch\"{o}n, in \textit{CFN Lectures on Functional Nanostructures}, Eds. K.
Busch \textit{et al.}, Lecture Notes in Physics \textbf{658} (Springer), pp.
145-164, (2005); M. Braun, J. K\"{o}nig, and J. Martinek, Superl. Microstr.
\textbf{27}, 333 (2005).

\bibitem {Gorelik}L. Y. Gorelik, S. I. Kulinich, R. I. Shekhter, M. Jonson,
and V. M. Vinokur, Phys. Rev. Lett. \textbf{95}, 116806 (2005).

\bibitem {Rudzinski}W. Rudzi\'{n}ski, J. Barna\'{s}, R. \'{S}wirkowicz, and M.
Wilczy\'{n}ski, Phys. Rev. B \textbf{71}, 205307 (2005).

\bibitem {Pedersen}J. N. Pedersen, J. Q.Thomassen, and K. Flensberg, Phys.
Rev. B \textbf{72}, 045341 (2005).

\bibitem {Fransson}J. Fransson, Europhys. Lett. \textbf{70}, 796 (2005).

\bibitem {Braig}S. Braig and P. W. Brouwer, Phys. Rev. B \textbf{71}, 195324 (2005).

\bibitem {Wetzels}W. Wetzels, G. E. W. Bauer, and M. Grifoni, Phys. Rev. B
\textbf{72}, 020407(R) (2005).

\bibitem {Parcollet}O. Parcollet, and X. Waintal, Phys. Rev. B \textbf{73},
144420 (2006).

\bibitem {Mu}H.-F. Mu, G. Su, and Q.-R. Zheng, Phys. Rev. B \textbf{73},
054414 (2006).

\bibitem {Grabert}H. Grabert and M. H. Devoret (Eds.), \textit{Single Charge
Tunneling}, (Plenum Press, New York, 1992).

\bibitem {Brataas}A. Brataas, Yu. V. Nazarov, and G. E. W. Bauer, Phys. Rev.
Lett. \textbf{84}, 2481 (2000); A. Brataas, Y. V. Nazarov, and G. E. W. Bauer,
Eur. Phys. J. B \textbf{22}, 99 (2001); A. Brataas, G. E. W. Bauer and P. J.
Kelly, Phys. Rep. \textbf{427}, 157 (2006).

\bibitem {Stiles}M. D. Stiles and A. Zangwill, Phys. Rev. B \textbf{66},
014407 (2002).

\bibitem {Tulapurkar}A. A. Tulapurkar, Y. Suzuki, A. Fukushima, H. Kubota, H.
Maehara, K. Tsunekawa, D. D. Djayaprawira, N. Watanabe, and S. Yuasa, Nature
\textbf{438}, 339 (2005).

\bibitem {Cottet}A. Cottet, T. Kontos, W. Belzig, C. Sch\"{o}nenberger, and C.
Bruder, Europhys. Lett. \textbf{74}, 320 (2006).

\bibitem {ZhangXue}P. Zhang, Q-K. Xue, Y. Wang and X. C. Xie, Phys. Rev. Lett.
\textbf{89}, 286803 (2002).

\bibitem {Martinek}J. Martinek, Y. Utsumi, H. Imamura, J. Barna\'{s}, S.
Maekawa, J. K\"{o}nig, and G. Sch\"{o}n, Phys. Rev. Lett. \textbf{91}, 127203 (2003).

\bibitem {Lopez}R. L\'{o}pez, and D. S\'{a}nchez, Phys. Rev. Lett.
\textbf{90}, 116602 (2003).

\bibitem {Stilesxc}M. D. Stiles, Nanomagnetism: Ultrathin Films, Multilayers
and Nanostructures (Contemporary Concepts of Condensed Matter Science, Vol 1),
eds. D. Mills and J. A. C. Bland, New York: Elsevier, 2006, pp 51-77.

\bibitem {Yuasa}S.Yuasa, T. Nagahama, and Y. Suzuki, Science \textbf{297}, 234 (2002).

\bibitem {Man}H. T. Man, I. J. W. Wever, and A. F. Morpurgo, Phys. Rev. B
\textbf{73}, 241401 (2006).

\bibitem {Waintal}X. Waintal, E. B. Myers, P. W. Brouwer, and D. C. Ralph,
Phys. Rev. B \textbf{62}, 12317 (2000).

\bibitem {Brataasrev}A. Brataas, G. E. W. Bauer, and P. J. Kelly, Phys. Rep.
\textbf{427}, 157 (2006).

\bibitem {Slonczewski}J. C. Slonczewski, Phys. Rev. B. \textbf{39}, 6995 (1989).

\bibitem {Dani}D. Huertas-Hernando, Yu. V. Nazarov, and W. Belzig, Phys. Rev.
Lett. \textbf{88}, 047003 (2002).

\bibitem {Xia}K. Xia, P. J. Kelly, G. E. W. Bauer, A. Brataas, and I. Turek,
Phys. Rev. B \textbf{65}, 220401(R) (2002).

\bibitem {Matveev}K. A. Matveev, Zh. Eksp. Teor. Fiz. \textbf{99}, 1598 (1991).

\bibitem {Glazman}L. I. Glazman, and K. A. Matveev, Zh. Eksp. Teor. Fiz.
\textbf{98}, 1834 (1990).

\bibitem {Mahan}G. D. Mahan, \textit{Many particle physics, }(Plenum Press,
New York, 1981).

\bibitem {Zwierzycki}G. E. W. Bauer, A. Brataas, Y. Tserkovnyak, B. I.
Halperin, M. Zwierzycki, and P. J. Kelly, Phys. Rev. Lett. \textbf{92}, 126601 (2004)

\bibitem {Kulik}I. O. Kulik and R. I. Shekhter, Zh. Eksp. Teor. Fiz.
\textbf{68}, 623 (1975).

\bibitem {Landau}L. D. Landau, and E. M. Lifschitz, Quantum Mechanics
(Non-relativistic theory), 3rd ed. (Pergamon Press, 1977).
\end{thebibliography}
\end{document}